# Spectrum Sharing Policy in the Asia-Pacific Region

Zhiyong Feng, Zhiqing Wei

**Abstract:** In this chapter, we investigate the spectrum measurement results in Asia-Pacific region. Then the spectrum sharing policy in the Asia-Pacific region is reviewed in details, where the national projects and strategies on spectrum refarming and spectrum sharing in China, Japan, Singapore, India, Korea and Australia are investigated. Then we introduce the spectrum sharing test-bed that is developed in China, which is a cognitive radio enabled TD-LTE test-bed utilizing TVWS. This chapter provides a brief introduction of the spectrum sharing mechanism and policy of Asia-Pacific region.

Radio spectrum resources play fundamental important roles in wireless communication systems. The rapid growing demand for wireless communication services and the inefficient spectrum allocation methods result in the scarcity of spectrum resources, which greatly hinders the development of future wireless communication systems [1][2].

For example, in China, the state radio regulatory commission of China (SRRC) divides the entire available magnetic spectrum resources into many frequency sections, and assigns them for different license services such as broadcast TV, cellular networks for exclusive use. Such fixed spectrum allocation approach ensures that wireless applications and devices don't cause harmful interference with each other. However, it will result in inefficient use of current radio spectrum. Some bands heavily occupied by busy radio services while other bands are rarely used. There are already great difficulties to find unassigned spectrum for new broadband wireless communication services such as time division long term evolution (TD-LTE).

One of the most promising solutions to overcome this problem is cognitive radio (CR). A CR device has the ability to identify and access an unoccupied spectrum band for temporarily usage. Therefore, CR is viewed as a technology to overcome the current inefficient usage of radio spectrum resources [3][4].

Due to its importance, a lot of research funds have been invested to develop cognitive radio technology. National research programs have already been founded to support CR technology in China such as the Major State Basic Research Development Program of China (973 Program), the National High Technology Research and Development Program of China (863 Program) and the National Natural Science Foundation of China. Government policy regulators such as SRRC are considering modifying current spectrum allocation polices in order to enable dynamic spectrum access technologies. However, if they fail to fully understand the current spectrum occupancy patterns, the investments and efforts may not produce expected results.

In the rest of this chapter, we firstly introduce the spectrum measurement in Asian countries. Then the spectrum sharing policies are introduced. Then the main spectrum sharing technologies, such as spectrum sensing and geo-location database based spectrum sharing schemes are introduced. Finally, the spectrum sharing test-bed in China is briefly introduced.

# 1. Spectrum Measurement in Asia-Pacific Region

Spectrum occupancy survey is essential for spectrum management which provides the policy makers with necessary information about the frequency usage pattern of different services in different frequency bands. Until now, several measurement campaigns have already been conducted in USA, Singapore, Vietnam, and Germany [5][6][7][8]. All these studies show a common finding that a large portion of assigned spectrum resources are seldom used while some particular spectrum bands are overcrowded. In this chapter, we survey the spectrum occupancies in Asia-Pacific region.

## 1.1 China

The spectrum occupancies in Beijing are measured by our team. We have measured the spectrum band of 440 MHz to 2700 MHz for two weeks in Beijing. This measurement not only fills the gap of current lack of knowledge about radio spectrum usage pattern in Beijing, but also finds out the frequency bands which are suitable for future dynamic spectrum access products such as CR devices. Our measurement consists of two parts, i.e., fixed measurement and mobile measurement, which are introduced in the following sections.

### 1.1.1 Spectrum measurement

#### 1.1.1.1 Fixed measurement

The fixed measurement is adopted to measure the spectrum band of 440 MHz to 2700MHz. Our fixed measurement was taken on the roof top of a 30-storey building near the central business district (CBD) of Beijing. There are no higher buildings surrounding the measurement site, which enables us to measure the radio activity accurately. The equipment includes omni-directional broadband antenna BOGER DA753G which has a frequency range of 75 MHz~3 GHz, Agilent high performance spectrum analyzer N9030A with a dynamic range from -154 dBm to 30 dBm and a laptop computer. As illustrated in Fig. 1 (left), the antenna is installed on the roof top and connected to the spectrum analyzer by a low loss cable. The spectrum analyzer is controlled by a laptop computer which is used for setting the parameters and saving measurement data. Both the spectrum analyzer and the laptop were kept in an indoor metal box as shown in Fig. 1 (right). The resolution bandwidth (RBW) of the spectrum analyzer is set as 200 KHz. The measurement started at June 2012 in Beijing and lasted seven continuous days. In total, about 100 billion data samples were collected.

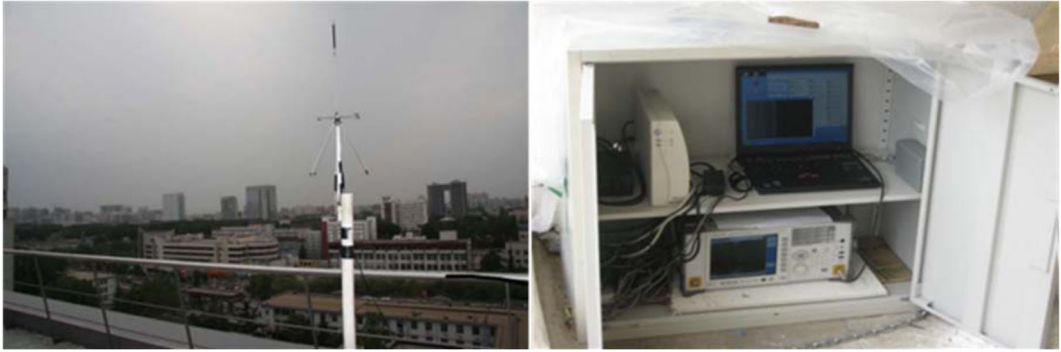

Fig. 1. Equipment and measurement environment of fixed measurement

## 1.1.1.2 Mobile measurement

The instruments utilized in mobile measurement and their connections are mostly the same as fixed measurement, as illustrated in Fig. 2 (left). It is worthwhile to mention that a GPS is connected to the computer so that the geographical coordinates of the data samples are recorded. The GPS used is Garmin 72HGPS which has a location precision of 5m. To ensure that the mobile measurement captures the actual spectrum occupancies in Beijing, a proper route is crucial. Therefore the following three principles are adopted for the selection of measurement route.

1) The route should be a helix circling the TV transmitters. In this way, the distance from the TV transmitter changes gradually, so that the relationship between radiation strength and distance can be captured.

2) To avoid the distortion introduced by Doppler effect, the speed of the vehicle should be kept low. Additionally, the performance of the instruments may be affected by the turbulence at high speed. Therefore, we maintained a speed around 20Km/h in the measurement.

3) To make the measurement convincing, the route should cover all typical places including business district, residents living quarters, open outdoor areas, etc.

The selected measurement route is plotted in Fig. 2 (right) with different colors denoting different kinds of places. In the map, yellow stands for residential areas, purple stands for business districts and white marks the open areas with sparse buildings. The measurement was conducted from 9 am to 5 pm in a weekday.

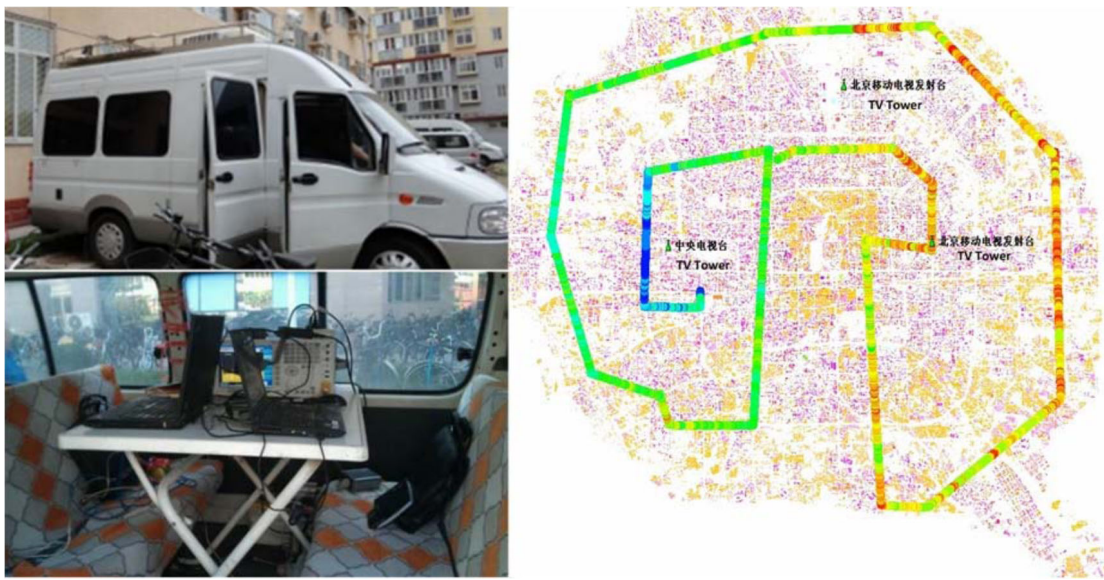

Fig. 2. Equipment and route of mobile measurement

## 1.1.2 The measurement results

### 1.1.2.1 Spectrum measurement of 440 MHz to 2700 MHz

The spectrum band from 450 to 470 MHz is authorized as radio navigation, radio location and land mobile services such as interphone. The average spectrum occupancy is observed as 0.18. A temporary usage pattern can be observed for part of the frequency channels in this band, which mainly are interphone services, resulting in relative low spectrum occupancy. In Fig. 3, spectrum measurement results in Beijing are provided.

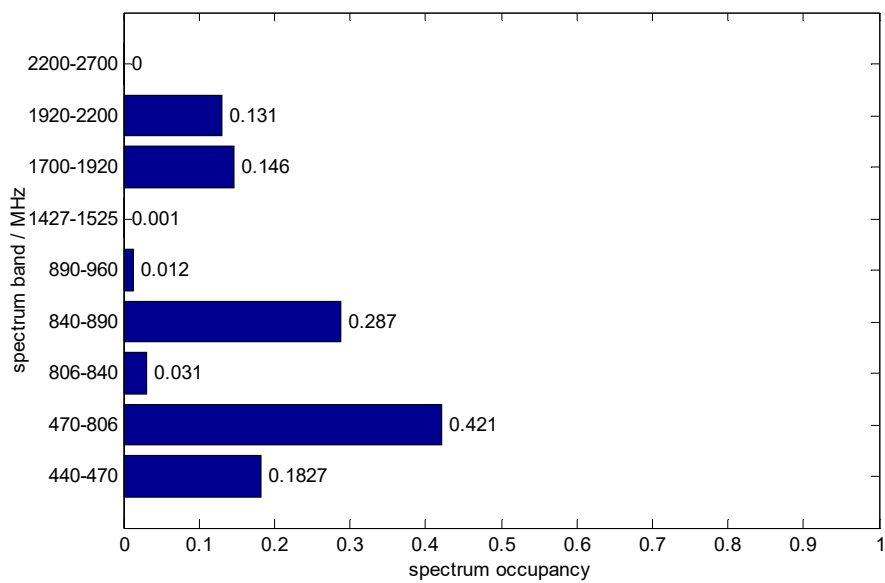

Fig. 3. Spectrum occupancy in Beijing

The spectrum band of 470 to 806 MHz is allocated to the broadcasting services, which are analogue Chinese TV service PAL-DK, and Chinese digital TV service DTMB. The average spectrum occupancy of spectrum band of 470 to 806 MHz is about 0.42. As we can see, not all of TV stations work 24 hours a day. Some TV channels are closed from 00:10 to 5:50.

The spectrum band of 880 to 960 MHz is assigned to the mobile cellular services (GSM900). The average spectrum occupancy of this spectrum band is 45.52%. Note that the spectrum occupancy patterns of uplink and downlink of GSM 900 are very different. Similar signal occupancy pattern are observed in the GSM1800 and 3G services. This may result from that the transmitting power of uplink for cellular systems is relative lower than the transmit power of downlink. And the uplinks are silent if there is no active communication. The TD-SCDMA band A (1880~1900 MHz) were not detected in our experiment, and its service only existed in band B (2010~2025 MHz). CDMA2000 signals were not detected during the 2 week periods. The average spectrum occupancy of 1700 MHz to 1920 MHz is about 14.6%. The average spectrum occupancy of 1920MHz to 2200MHz is about 13.1%.

The spectrum occupancy of industrial, scientific & medical band (ISM band), which is range from 2400 to 2500MHz, appears completely unused during 2 week observation. High likelihood signal occupancy pattern is found in the radar frequency band from 2500 to 2700 MHz, TD-SCDMA candidate band C 2300~2400 MHz and 1427~1525 MHz band for point to multipoint microwave communication system, which is currently utilized by military services. The occupancy estimations of ISM and radar bands may not be the real situation of the systems such as WLAN, Bluetooth. Since the measurement site is on the rooftop of a 30 storey (115meters high) building, the transmitter of the ISM signals may not reach to the measurement point. Besides, ISM signals may not penetrate through walls. Radar signals may need special detection methods and equipment since its pulse is too short to be detected for common frequency sweep mode receivers, resulting in a very low probability to be captured.

The summary of average spectrum occupancy in Beijing city is presented in Fig. 3. The results suggest the sparse usage characteristics both in time and frequency domains. The spectrum occupancy rate in average is 13.5% in Beijing for the 450~2700 MHz frequency bandwidth. That means nearly 86.5% of allocated spectrum was unused for this period.

## 1.1.2.2 Spectrum measurement of TV band

We conduct a fixed and a mobile measurement of the 470~806 MHz bands in Beijing. The frequency, time and space domain and the specific TV standards are all considered in our analysis. Similar with previous measurement, the fixed measurement and mobile measurement are both conducted.

In China, the spectrum band originally allocated for terrestrial TV broadcasting service is 470~806 MHz. Due to historical reasons, the 566 MHz~606 MHz band is reallocated to trunking communication service, which is omitted in our analysis. Thus the TV band takes up to 296 MHz bandwidth. Currently both analog television (ATV) and digital television (DTV) are utilized in Beijing. The channel bandwidth is 8 MHz for both DTV and ATV. Therefore there are totally 37 channels.

The average, maximum and minimum received power versus frequency during the 7-day measurement period is plotted in Fig. 4. It can be observed that the spectrum utilization can be coarsely classified into three cases. In the first case, the maximum power is close to the average power and both of them are very high, which means that these channels are occupied almost all the time, such as 750~758 MHz. In the second case, both the maximum and the average power are high. However, certain difference exists between them, which suggests these frequency points are only partially used in time, such as 478~486 MHz. In the last case, the maximum power is high while the average power is close to the noise floor, such as 700~730 MHz. Since TV broadcasting usually functions stably over a long period of time, it is very likely these channels are utilized by wireless microphone or interphone.

The water fall map of the spectrum band is depicted in lower subfigure of Fig. 3. It can be observed that some spectrum bands are not utilized from 00:00~06:00. To check whether spectrum utilization has periodicity over time, the spectrum occupancy versus time in the 7-day measurement is plotted in Fig. 4.

It is very obvious that occupancy of the entire spectrum band shows strong periodicity and the period is one day. As illustrated in Fig. 5, the lowest occupancy appears in 00:00~06:00, which can be explained by the fact that some TV transmitters are shut down at midnight due to small number of audience. The occupancy is almost identical for each day in a week, which proves the static property of the TV channels. Ripples in the occupancy curve can be explained by measurement errors and illegal usage. It needs to be noted that the high utilization period of TV band coincides with the busy period of public communication system such as CDMA and WLAN. Even though time domain spectrum opportunity is sufficient in 00:00~06:00, The spectrum demand may not exists in this period, which is an interesting paradox.

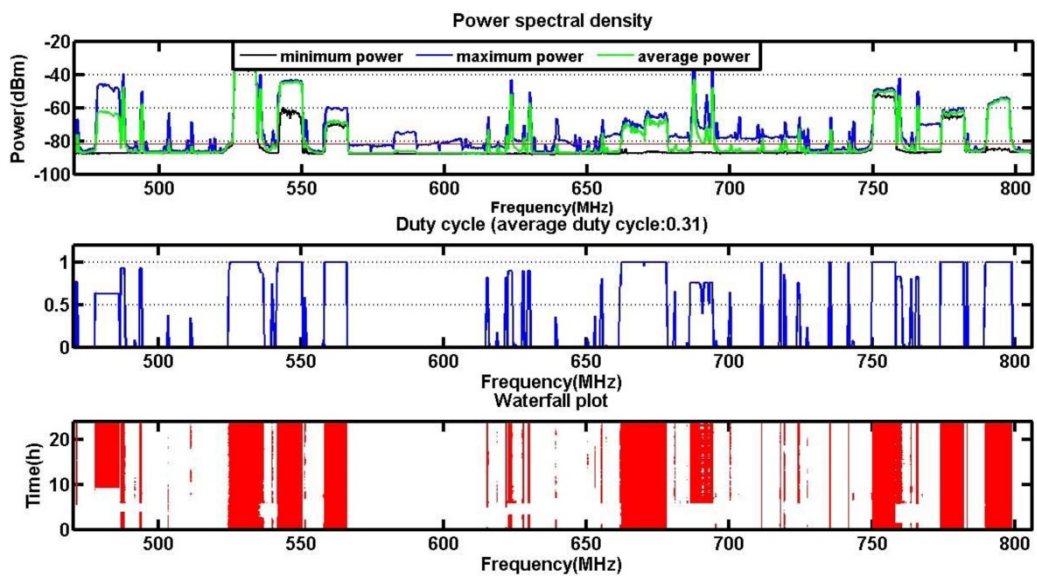

Fig. 4. The utilization of TV band

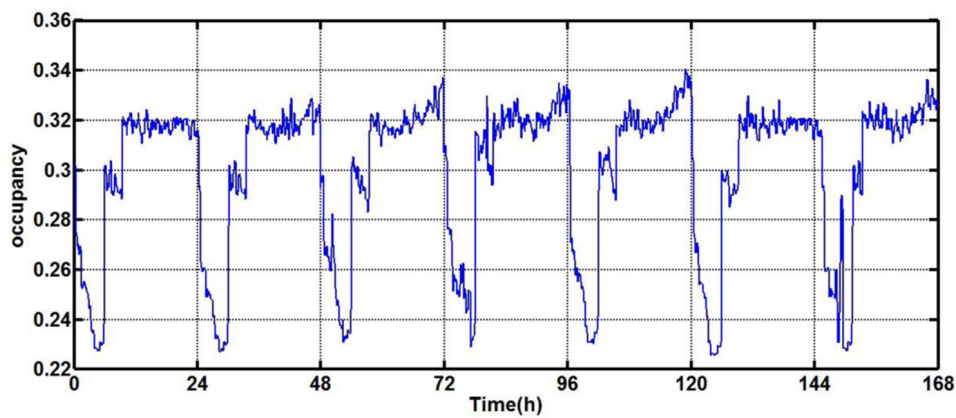

Fig. 5. Spectrum occupancy versus time in the 7-day measurement

## 1.1.1.3 Suggestions on spectrum innovation in TV band of China

Based on the analysis of both fixed and mobile measurement above, we would like to provide our suggestions to spectrum innovation in the TV band of China.

Our first suggestion is that China should accelerate the transition from ATV to DTV. The advantage of DTV is four fold. First of all, the DTMB standard developed by China can provide a data rate as high as 30Mbps using OFDM, which is sufficient for HDTV. Therefore, the watching experience will be greatly enhanced compared to ATV. Secondly, the receiver design and out-of-band performance of DTMB is far better than PAL-DK, which makes it robust to adjacent channel interference. Replacing ATV with DTV will eliminate the hidden occupancy problem, resulting in fewer requirements on the device using this band. Thirdly, Chinese DTV standard has its unique single frequency network (SFN) technology. Traditionally, to cover a huge area, the same TV program is broadcasted using different channels to avoid interference among transmitters. In SFN, one TV program can be synchronously broadcasted across the entire area using only one channel. Thus spectrum efficiency can be improved and more channels can be relieved for reallocation. Lastly, multiple TV programs can be carried by one DTV channel and the channels needed for TV broadcasting can be further reduced. The transition to DTV is an international trend which has been accomplished in many countries. China began its transition in 2008 and is very likely to accomplish the process in 2020.

We believe spectrum reallocation should be the main spectrum innovation strategy in the TV band of China. As the measurement indicates, the overall occupancy of TV band is merely 0.38, which is very low. Therefore the vacant band can be reallocated to other services. However, the spectrum allocation for TV varies in different area of the country. To leave a universal band for other service, national wide adjustment of TV frequency may be required. Moreover, the spectrum occupancies show that the time domain spectrum opportunity in the midnight and space domain spectrum opportunity in the suburb of city exist. However, there are difficulties to use these spectrum opportunities. Besides, indoor spectrum opportunities exist for short distance communication technologies such as femtocell. However, further measurement is needed to justify the possibility to implement femtocell in TV white space.

## 1.2 Other countries or regions

Contreras *et al.* in [11] carried out the spectrum occupancy measurements in three locations of Kanto area in Japan. The spectrum band of 90 MHz to 3 GHz is measured. It is found that only 6.9% of the spectrum between 90 MHz and 3 GHz is used for more than 10% of time, which indicates great potential for deploying spectrum sharing systems in Japan.

The spectrum occupancies of different sub-bands are provided by [11], which is presented in Table 1. It is noted that the spectrum occupancy of 810~958 MHz is highest, which is occupied by cellular networks. On the contrary, the spectrum bands which are occupied by radar, TV broadcasting system, etc. are relatively idle. Overall, the spectrum measurement in Japan has demonstrated great potential for deploying spectrum sharing systems in Japan all along the evaluated spectrum from 90 MHz to 3 GHz [11].

Table 1. Spectrum occupancies in Japan [11]

| Subband | Bandwidth | Occupancy | Services |
| --- | --- | --- | --- |
| 90~108 MHz | 18 MHz | 23.58% | TV broadcast (until 07/11) |
| 108~170 MHz | 62 MHz | 12.91% | Miscellaneous Communications |
| 170~222 MHz | 52 MHz | 17.46% | TV broadcast (until 07/11) |
| 222~470 MHz | 248 MHz | 0.63% | Miscellaneous Communications |
| 470~810 MHz | 340 MHz | 13.9% | TV (until 07/12), Radio microphones |
| 810~958 MHz | 148 MHz | 28.55% | Cellular networks, disaster prevention, RFID |
| 958~1710 MHz | 752 MHz | 0.09% | Miscellaneous services |
| 1710~2300 MHz | 590 MHz | 5.9% | IMT 2K, spatial applications |
| 2300~3000 MHz | 700 MHz | 0.018% | Radar, IMT2K, ISM, public communications |

Jayavalan *et al.* in [12] investigated the spectrum occupancy of cellular and TV broadcasting spectrum bands in Malaysia, which is provided in Table 2. It is found that some TV spectrum bands are not utilized in Malaysia due to cross-border frequency coordination because the country is closely located with Singapore, Thailand and Indonesia. Specifically, TV broadcasting in VHF and UHF bands has the average duty cycles of 11% and 13% [12]. Besides, Jayavalan *et al.* found that most of the allocated TV spectrum bands are below 15% utilization. Thus their work has provided the motivation for spectrum sharing in Malaysia, which can bring economic and social welfare to the country.

Table 2. Spectrum occupancies in Malaysia [12]

| Service | Frequency Range (MHz) | Bandwidth (MHz) | Average Duty Cycle (%) |
| --- | --- | --- | --- |
| GSM 900 | 880~960 | 80 | 35.31 |
| GSM 1800 | 1710~1880 | 170 | 9.59 |
| 3G (IMT-2000) | 1885~2200 | 315 | 26.08 |
| VHF TV | 174~230 | 56 | 10.92 |
| UHF TV | 470~798 | 328 | 13.36 |

# 2. Spectrum Sharing Policy in the Asia-Pacific Region

## 2.1 China

According to the prediction of ITU, the spectrum demand of international mobile telecommunications (IMT) in 2020 is 1340~1960 MHz. In China, the SRRC estimates that spectrum demand in 2020 is 1490~1810 MHz. However, there is still 1000MHz shortage of spectrum for China.

In this situation, the spectrum refarming, spectrum sharing, etc. are proposed to improve the spectrum utilization. In China, government policy regulators such as ministry of industry and information technology (MIIT), SRRC, etc. are considering modifying current spectrum allocation polices in order to enable dynamic spectrum access technologies. China government has already released the regulations that allow the spectrum sharing among the radio access technologies (RATs) within an operator. In 2016, MIIT of China firstly carried out spectrum audit, which aims to evaluate the spectrum usage to enhance the radio spectrum resource regulation. Through this work, Chinese government obtains the actual use of key spectrum bands of seven public mobile communication systems consisting of 25 spectrum bands. Thus Chinese government has enhanced the monitoring level of spectrum, which paves the road for spectrum refarming and spectrum sharing.

Besides, to satisfy the requirements of future mobile broadband systems in 5G, such as ultra-high traffic and ultra-high data rates, more spectrum and wider bandwidth are needed for realizing the performance. Therefore, the innovative spectrum utilization methods should be further investigated to discover the available frequency of IMT under horizontal or vertical spectrum sharing systems. Spectrum share means dynamic and optimized allocation of multi-RAT spectrum resource. It considers some key factors such as deployment scenario, network load and user experience. Spectrum efficiency improvement and interference control are benefited from optimized dynamic spectrum allocation and management among different networks or systems. Besides, self-adaption functions of autonomous access network and handover between networks can be realized. Furthermore, air interface efficiency and coverage are enhanced for the efficient, dynamic and flexible spectrum utilization. Therefore integrated frequency utilization efficiency can be increased. When using joint spectrum, network components are likely to play a balancing role.

Spectrum sharing technology can be classified by the application scenarios, which include intra-operator inter-RAT spectrum sharing, inter-operator spectrum sharing, spectrum sharing in unlicensed band, and spectrum sharing in secondary access. The implementation choices for spectrum share are various, which include independent control node, database based control etc. The specific functionalities of spectrum share contain multiple priority spectrum allocation, interference coordination and so on. In the following sections, we introduce the development of spectrum sharing technologies in China.

## 2.1.1 Spectrum refarming

In American, Finland, etc., the spectrum bands of 2G mobile communication systems are reallocated to the 4G mobile communication systems. In China, there are also some cases of spectrum refarming. In 1998, the personal handy-phone system (PHS) entered Chinese market. Until October, 2006, the number of users of PHS system in China is 93 million. However, in October, 2008, MIIT of China announced that the spectrum band of PHS, namely, 1900~1920 MHz is planned to be reallocated to time division-synchronous code division multiple access (TD-SCDMA) system, which is a third generation mobile communication system designed by China. In January 2011, PHS system was beginning to be officially withdrawn from the market of China. In October, 2014, the base stations of PHS are closed in China.

TD-SCDMA has inherited the spectrum band of PHS system in China. However, when the TD-SCDMA starts to commercially operate, the 4G mobile communication system is already on the way. Besides, The time division long term evolution (TD-LTE) and TD-SCDMA are both operated by China Mobile Communications Corporation (China Mobile). Hence TD-SCDMA is in an awkward situation. In 2013, China Mobile determined that the voice service of 4G will fall back to GSM system other than TD-SCDMA system. In 2014, China Mobile announced that the investment to TD-SCDMA is stopped. And the users of TD-SCDMA will move to TD-LTE, which means that the TD-SCDMA network will be naturally withdrawn in the future. Actually, China Mobile has already started some trials to refarm the spectrum of TD-SCDMA. In 2015, China Mobile has allocated the spectrum band A (1880~1920 MHz) of TD-SCDMA to TD-LTE, such that the data rate of TD-LTE can be improved to 79.8 Mbps. Besides, China Telecom and China Unicom start to refarm the spectrum of CDMA 800MHz and GSM 900MHz [13].

## 2.1.2 Spectrum sharing

In China, the mobile Internet plays an important role in economic development and everyone's life. Driven by the devices such as smart phones, tablet PC and the services such as social networking, streaming media and online game, the amount of mobile traffic is increasing exponentially, which gives birth to the fifth generation mobile networks (5G). One of the QoS (Quality of Service) requirements of 5G is the 1000X capacity improvement compared with 4G. With such large capacity demand in the future mobile communication system, the spectrum refarm cannot solve the shortage of spectrum. Hence the spectrum sharing is regarded as the long term solution in the 5G era.

Recently, with the emergence of sharing economy or collaborative consumption, the spectrum sharing is also becoming more and more popular. Similar with the business model of Uber and DiDi, the government believes that spectrum sharing can stimulate the economy. In 2013, Broadband China Strategy was proposed by the government of China, where optimized spectrum planning is listed as one of the strategies.

In the spectrum planning of Broadband China Strategy, the government of China has regarded dynamic spectrum allocation as the promising approach to improve the utilization of spectrum resources. To realize this goal, the interference mitigation technology and equipment are required to be addressed to enable the spectrum sharing among different wireless services.

Meanwhile, the spectrum regulation for multiple wireless devices operating on the public spectrum is essential, which aims to maintain the order of spectrum utilization among the spectrum sharing wireless networks and devices.

The TV white spaces (TVWSs), 2.3 GHz and 3.5 GHz, etc. are commonly regarded as the promising spectrum bands that can be shared by other wireless systems. In the Broadband China Strategy, the spectrum band of 2300~2400 MHz, which is originally allocated to radar system, was planned to be shared with wireless communication systems. The spectrum bands that are utilized locally, such as spectrum bands of TV and radar, are commonly regarded as the ideal candidate spectrum sharing bands. However, the TV white spaces in China are hard to be shared because of policy issues, although the sharing of TV white spaces is widely studied in Chinese academia.

## 2.1.2.1 National projects of cognitive radio

The government of China has supported a lot of national projects studying cognitive and spectrum sharing, etc. In 2005, National High-tech R&D Program (863 Program) began to support the study of cognitive radio. In 2008, National Natural Science Foundation of China establishes a group of projects studying cognitive radio. Beijing University of Posts and Telecommunications (BUPT), Tsinghua University, PLA University of Science and Technology, etc. were supported by this group of projects.

Specifically, in 2008, BUPT was supported by the Major State Basic Research Development Program of China (973 Program) to study the architecture of cognitive radio networks, the multi-domain cognition theory and method, autonomous resource management and control schemes of cognitive radio networks, etc. The architecture of cognitive radio network is proposed in Fig. 6. The architecture consists of four modules.

1) End-to-end goals management. This module consists of two aspects. Namely, the end-to-end goals and cognitive specification language. The cognitive information collection and autonomous resource management module are guided by the end-to-end goals. The cognitive specification language maintains the connection among different modules.
2) Cognition coordinator. Cognition coordinator is mainly used to set the filter to obtain cognitive information from multi-domain environment. Then cognition coordinator establishes the precise mapping between multi-domain environment and cognitive information. Finally, cognition coordinator extracts the relevant cognitive information.
3) Self-organization coordinator. With the goals of end-to-end utility, the autonomous resource management module generates the optimization strategy based on the collected cognitive information.
4) Reconfiguration management. Based on the optimization decision, the related reconfiguration model is established to execute the reconfiguration command and reconstruct the network parameters.

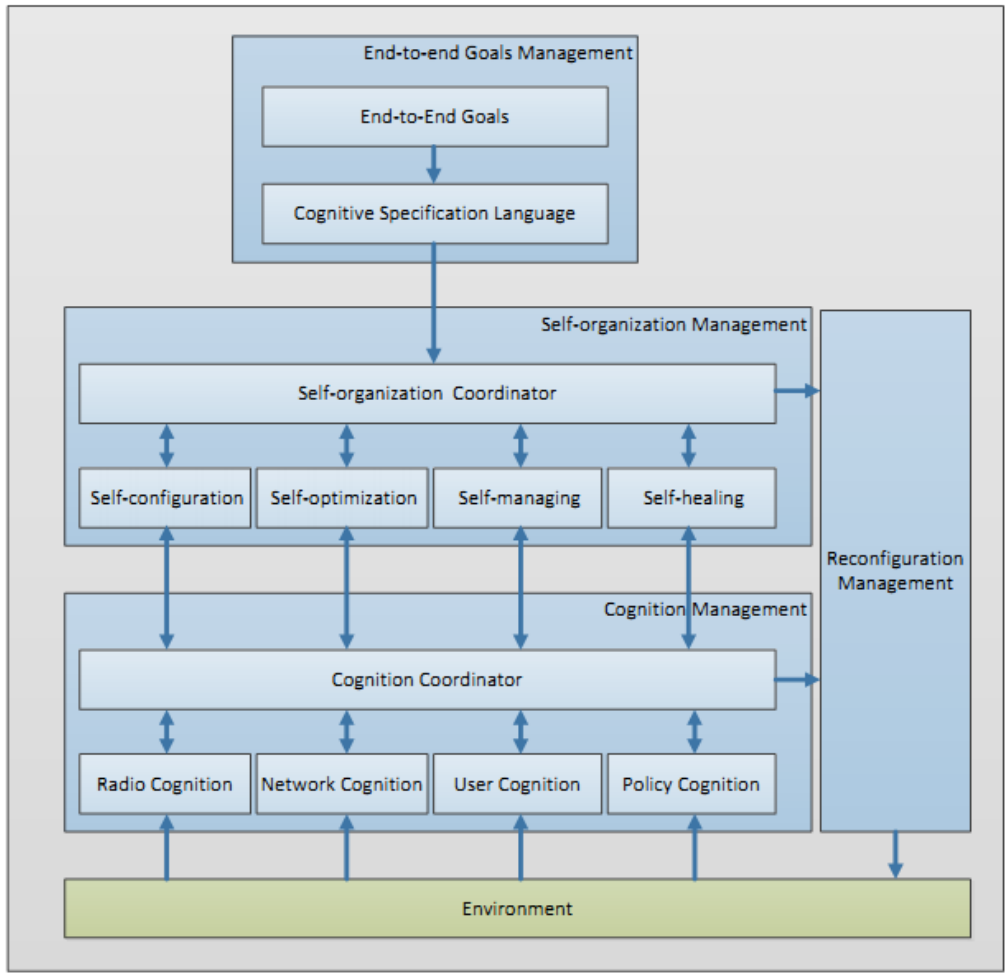

Fig. 6. The architecture of cognitive radio proposed by China

In the architecture of cognitive network, the entire architecture is driven by end-to-end performance. And the learning function is realized through the interaction between the theoretical model and the modules. In this model, the cognitive information flows from the cognitive management module to the end-to-end module, the self-organization module, and the reconfiguration module, which is illustrated in the Fig. 7.

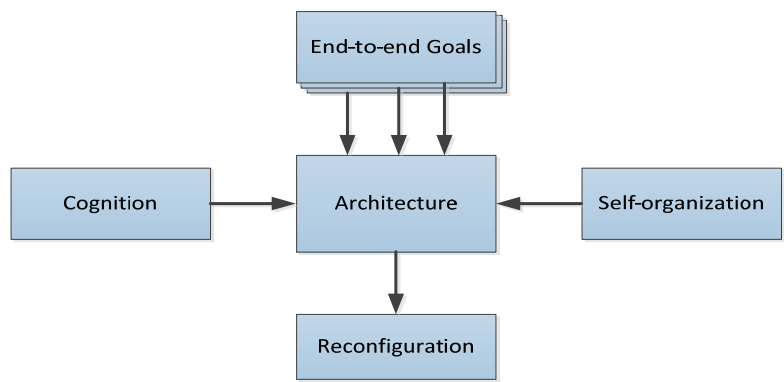

Fig. 7. The main modules in the architecture of cognitive radio.

Besides BUPT, University of Electronic Science and Technology of China (UESTC) was also supported by 863 Program. UESTC had designed the architecture of cognitive radio networks and developed the test bed of cognitive radio networks with the support of 863 program.

## 2.1.2.2 Cognitive radio networks in China

It is noted that in China, the cognitive TD-LTE system is proposed by BUPT, which can be regarded as one of the precursors of LTE-U. In LTE-U, LTE system shares the unlicensed band. Cellular network is the pillar of telecommunication industry. Thus utilizing cognitive technology to solve spectrum usage in cellular network is of great importance. To improve the spectrum efficiency of cellular system, the following requirements must be satisfied.

1) Accurate and efficient vacant spectrum awareness.
2) Dynamic spectrum management.
3) Flexible and adaptive transmission and dynamic spectrum utilization.

Notice that TDD mode can be operated in unpaired spectrums, whereas FDD requires paired spectrums. Thus TDD offers more flexibility in spectrum allocation. With the assistance of cognitive radio, the capacity of TD-LTE system can be improved.

The cognitive TD-LTE system which coexists with TV broadcast services enables high efficiency of spectrum utilization. Based on the field test of the spectrum band allocated to broadcast TV services, 698~806 MHz is selected for the cognitive TD-LTE system. However, to guarantee the control signaling and basis data transmission, the dedicated spectrum is also necessary. The dedicated spectrum is allocated to a certain cognitive cellular system to guarantee cognitive information transmission and basic data transmission. Different dedicated spectrum bands are allocated to multiple cognitive cellular systems to work together without interference to each other, which are illustrated as follows. Hence it is challenging to design dynamic spectrum management for the cognitive cellular system with statically utilized dedicated frequency and dynamically utilized cognitive frequency, which is illustrated in the following figure.

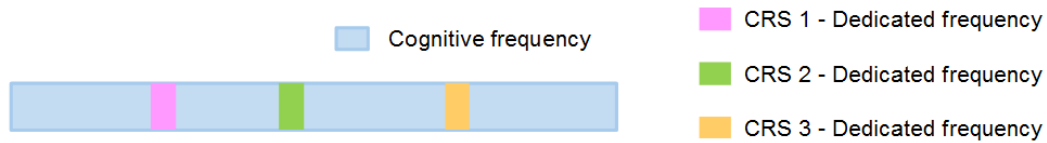

Fig. 8. The spectrum allocation in cognitive radio.

Besides, as illustrated in Fig. 9, two-level spectrum management architecture is proposed to enable the spectrum regulation among TD-LTE and TV broadcast services, which are global layer and local layer.

1) Global layer: As illustrated in the following figure, the advanced spectrum management (ASM) server is operated in the global layer. The ASM is responsible for the inter-cell spectrum management, which is a large time granularity spectrum planning.
2) Local layer: In practice, the spectrum occupancies are varying in the space and time dimensions. Hence the local layer spectrum management is also needed. In the local layer spectrum management, the combination of cognitive database and spectrum sensing is essential to capture the spatial and temporal spectrum fluctuations and coordinate the possible interference between TD-LTE and TV broadcasting networks.

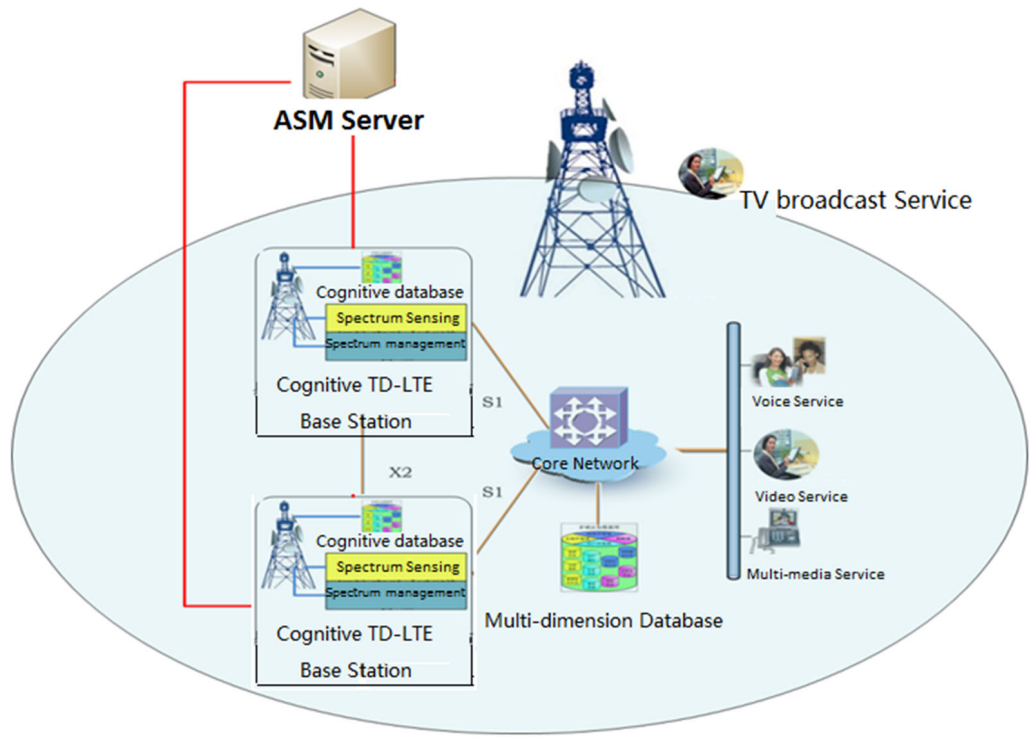

Fig. 9 The two-level spectrum management architecture.

## 2.1.3 Policy and regulation challenges of spectrum sharing in china

The major difference of spectrum policy between China and western countries is that spectrum bands are not allocated by auction. Thus the operators in China, such as China Mobile, China Unicom and China Telecom do not need to pay for the dedicated spectrum, which is different from other countries. In the United States, Britain or Germany, etc., the spectrum is always billion dollars asset. Hence in these countries, the spectrum sharing is highly motivated to improve the spectrum efficiency. When Qualcomm proposed the LTE-U and LAA technologies, the motivation was that the dedicated spectrum is expensive. When the operators want to improve the network capacity, the unlicensed spectrum is reasonable choice. Hence in China, in the operators' point of view, the motivation of spectrum sharing is not strong compared with other countries. Meanwhile, the sharing of TV white spaces is not well developed because of the policy issues.

However, on the other hand, Chinese government has strong incentive to push spectrum sharing forward because the sharing of spectrum can stimulate the industry innovation and boost the economic development. The spectrum regulation entities such as MIIT, SRRC, etc, have launched a large number of regulations and projects to study the possibility to share the spectrum bands of TV and radar. The sharing of infrastructures and spectrum can save public resources and avoid the waste of resources, such as reputably building the base stations and low utilization of spectrum. Hence Chinese government has promoted to establish the China Tower company, which aims to share the base stations among Chinese operators. Therefore, with the vision of development, the spectrum sharing in China is challenging, but it is on the way.

## 2.2 Japan

The spectrum policy of Japan is administered by the radio regulatory commission, which is an independent agency under Japanese government [22]. The Ministry of Internal Communications (MIC) of Japan established a committee in 2004 to formulate policies related to spectrum regulation. In 2006, in order to share the spectrum including 700 MHz spectrum bands, MIC spent one year to solicit opinions in civil communities. The final decision is to retain the 10 MHz in the 60 MHz UHF spectrum band for the deployment of intelligent transport system (ITS) system.

Japan began to levy spectrum usage fees from 2005. In February 2014, Japan adopted the amendment on the Radio Law at the cabinet council. The main modification is to reduce the cost for operators paid for using the spectrum from the beginning of 2014 [23]. In the past, only television, radio stations and other public property units can get such benefit.

In Japan, with the innovation of wireless technologies such as small cells, code division multiple access (CDMA), and high speed packet access/evolution data only (HSPA/EVDO) [15], the spectrum utilization is significantly improved compared with traditional spectrum usage. However, the spectrum in Japan is nearly exhausted. Hence the emerging wireless technologies have great difficulties in finding new spectrum bands, which is similar to China. With the severe spectrum usage, the development of mobile Internet faces great challenge. However, the economy is boosted by the information and communication technology (ICT), where the mobile communication plays an important role. With the upgrade of different generations of wireless communication systems, the ideal frequency bands are allocated to the prior wireless communication system, which makes the spectrum resource scarce.. Besides, the mobile communication operators need to provide new services to attract consumers. Therefore the spectrum resource is also in short supply. Hence the mobile communication operators need to find the approach to efficiently use spectrum. In this situation, spectrum sharing is one of the promising technologies to improve the spectrum efficiently [15].

## 2.2.1 Cognitive radio architecture proposed by Japan

Spectrum sharing aims to use the vacant spectrum bands. In spectrum sharing, the vacant spectrum bands and vacant time slots of primary users can be exploited by secondary users. By sensing and accessing the vacant spectrum in spatial and temporal dimensions, secondary users can use adequate spectrum bands and improve the capacity of cognitive radio networks. In the implementation of spectrum sharing in cognitive radios, the cooperation among heterogeneous wireless networks is essential to coexist among them and improve the spectrum utilization.

In Japan, the Cognitive Wireless Cloud (CWC) was proposed by NICT [16], which is architecture of cognitive radio network. The cognitive radio systems which are cooperated with primary networks are called dynamic spectrum access network or cognitive wireless clouds [16]. In CWC, a network reconfiguration manager (NRM) installed in secondary networks collects the sensing data of secondary users. The sensing data includes data rate, delay, throughput and signal strength, etc. The NRM analyzes the collected sensing data and feedbacks the controlling information to secondary users, such that secondary users can access to the most appropriate wireless networks.

The CWC enables the radio equipment with multiple air interfaces to autonomously utilize the most appropriate wireless networks and spectrums by configure their own transmission parameters. The key functions of CWC are as follows [16].

- Collecting context information from the available wireless networks or spectrums.
- Storing context information in cloud, which can also be a database.
- Analyzing the available context information.
- Dynamically making spectrum access decisions that can satisfy the goals of users or operators as follows.
    - ✓ Maximization of capacity,
    - ✓ Minimization of interference,
    - ✓ Co-existence among wireless networks.
- Reconfiguring the wireless networks seamlessly.

## 2.2.2 White space sharing in Japan

It is worth noting that Japan is a small island country with a population of 126.81 million. And it is also close to Busan, Korea's major city. Therefore the geographic features of Japan are different compared with many countries that have multiple land borders with other countries. Large population and small area make Japan's spectrum resources particularly scarce. The government of Japan attaches great importance to the efficient use of spectrum.

In order to solve the shortage of spectrum resources, MIC released a series of related policies to improve spectrum utilization. The exploitation of white space is an important part. MIC announced the action plan for spectrum reuse in 2004. And the latest revision was completed in October 2015 [17]. For the implementation of the plan, the government of Japan conducts annual audits for spectrum. Through the investigation and evaluation of specific spectrum bands, they developed spectrum recovery programs, and use the frequency compensation mechanism to promote migration.

In July 2011, Japan completed the transition from analog TV to digital TV. Spectrum band of 710~722 MHz was released in July 24, 2006. Spectrum band 722~770 MHz was released in July 24, 2012, which was no longer used for the broadcasting service [18]. To solve the problem of how to use free spectrum, the MIC established the "radio broadband promotion committee" in November 2004. In March 2006, in order to share three frequency bands 90~108 MHz, 170~222 MHz and 710~770 MHz, the government formally collected external solicitation of specific spectrum applications [19].As illustrated in Fig. 10, TV white space (TVWS) devices must be separated from the protected area of TV channel by a distance based on their transmission power and antenna height. Typically, the protected area of TV channel is the area where the corresponding transmitter provides the required signal strength for acceptable TV service. According to [20], there are more than 20 TVWS channels available in central Tokyo in 2012. Overall, many areas in Japan will have at least 100 MHz TVWS. There are many wireless systems aiming to be operated in TVWS, such as sensor networks, wireless microphones, and public safety communication systems. The possibility of adopting geo-location database in spectrum sharing is being investigated by the TVWS working group established by MIC, which has so far investigated the feasibility of allowing small-area broadcasting systems, wireless microphones, sensor networks, and public safety communication systems to access TVWSs [21] (Table 3).

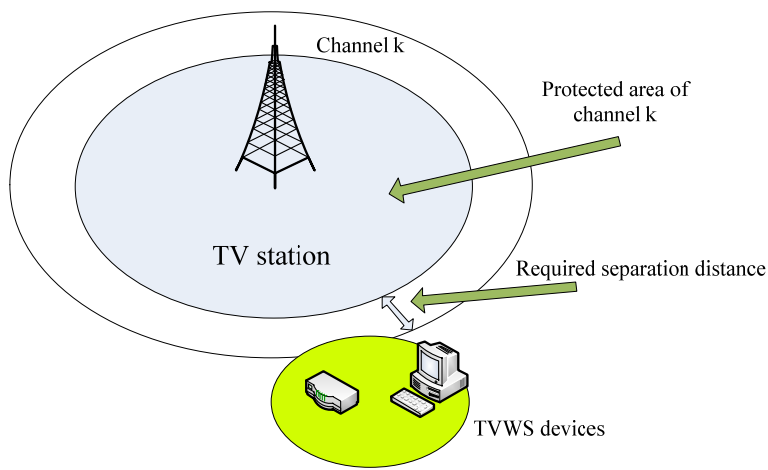

Fig. 10 Basic concept of TVWS device regulations [20].

Table 3  TV Channel allocation in six largest metropolitan areas [20-21]

| Channel | 13 | 14 | 15 | 16 | 17 | 18 | 19 | 20 | 21 | 22 |
|---|---|---|---|---|---|---|---|---|---|---|
| Tokyo |  |  |  | tv |  |  |  |  | tv | tv |
| Osaka | tv | tv | tv | tv | tv | tv |  |  |  |  |
| Nagoya | tv |  |  |  |  | tv | tv | tv | tv | tv |
| Hiroshima |  | tv | tv |  |  | tv | tv |  |  | tv |
| Miyazaki | tv | tv | tv | tv |  |  |  |  |  |  |
| Fukuoka |  |  |  |  |  |  |  |  |  | tv |
| 23 | 24 | 25 | 26 | 27 | 28 | 29 | 30 | 31 | 32 | 33~52 |
| tv | tv | tv | tv | tv | tv |  |  |  |  |  |
|  | tv |  |  |  |  |  |  |  |  |  |
| tv |  |  |  |  |  |  |  |  |  |  |
| tv |  |  |  |  |  |  |  |  |  |  |
|  |  |  |  |  |  |  |  |  |  |  |
|  |  |  | tv |  | tv |  | tv | tv | tv | tv (34) |

In Japan, the spectrum has already been allocated to a variety of services. However, although most of spectrum bands have been occupied, the actual utilization of spectrum is only 10% to 20%. In Japan, spectrum sharing also faces the political barriers, which is not a pure technical problem. Hence the realization of spectrum sharing still takes a long way to go.

## 2.3 Singapore

## 2.3.1 White space sharing in Singapore

Similar to Japan, due to its restricted area and large population, the government of Singapore is very concerned about the efficient use of spectrum. There are also many trial experiences to investigate the possibility to use spectrum efficiently. A Singapore survey pointed that the actual spectrum utilization was only 6.5%. Analog TV to digital television transition produces a large number of idle spectrum bands and the new spectrum utilization method, namely, spectrum sharing is proposed to improve spectrum efficiency [24].

TVWS is a good choice to be utilized to enrich available spectrum. 24 channels in the TV broadcasting bands will be available for TVWS operations in Singapore. The VHF spectrum (174 MHz and 230 MHz) and the UHF spectrum (470 MHz and 806 MHz) for TVWS operations were not fully utilized [26]. Such spectrum bands could potentially be redeployed for other wireless services. Info-communications development authority in Singapore (IDA) has the authority to plan the spectrum bands. The white space equipment operates without permission as long as it complies with IDA's technical specifications. The license-free approach allows users to explore a range of business models, which can reduce user's costs and stimulate innovative services.

As early as 2006, Singapore institute of information and communications (I2R) participated in the standardization of TVWS in IEEE [27]. In 2008, I2R proposed experimental prototype exploiting TVWS to FCC, which is the original development of spectrum sharing in Singapore. In 2009, IDA has finished Singapore's regulatory framework for TVWS deployment [28].

In 2011, IDA carried out the spectrum planning in white spaces and development of relevant regulatory framework. In 2012, Singapore set up a TVWS test working group. More than 40 locations were tested within two years and it is found that the use of TVWS has many obvious advantages. It is estimated that Singapore has TVWS of up to 180 MHz. Once the TVWS is open for use, the capacity of wireless networks can be boosted. IDA has developed a regulatory framework in November 2014 aiming to allocate the 180 MHz TVWS to unlicensed users.
In 2015, TVWS regulatory framework was further pushed forward. It allows other wireless networks to use TVWS to increase capacity to create broadband wireless networks. Singapore is the second country in the world after United States that agrees to access TVWS for unlicensed users [29]. The concept of dynamic spectrum access can optimize and enable the access to under-utilized spectrum, which can be used in many areas such as smart grid, wireless broadband and machine to machine communication. Moreover, the geo-location database approach will be a practical method for white space devices to access TVWS. Allowing unlicensed users accessing TVWS will motivate potential new technologies and business models, which will bring opportunities for small and medium sized enterprises. Although companies wish to use TVWS without license, they need to comply with the technical regulations proposed by IDA. Manufacturers and suppliers of TVWS equipment must register their devices in IDA, which is similar to the manufacturers and suppliers of mobile phones and communication equipment [30].

Further, Minister of Communications and Public Information in Singapore announced the latest development blueprint, namely, master plan 2025, to tackle the shortage of spectrum facing the great spectrum demand of mobile Internet [31].

## 2.3.2 National projects of cognitive radio in Singapore

Singapore has developed a white space testbed cognitive radio venue (CRAVE) to evaluate the promising technologies to improve spectrum utilization. The test locations of CRAVE are Malaysian coast, Indonesia coast, near broadcast tower, dense urban environment and in-building environment [32]. Through this program, the performance of WSD devices is evaluated.

## 2.3.3 Policy and regulation challenges of spectrum sharing in Singapore

Singapore is a country with small area of land and limited natural resources. Facing the growing number of elderly population and the process of urbanization, the high-tech products such as sensors, robots and wireless networks are widely deployed to build a smart nation that improves people's quality of life. It's also the common aspiration of Singapore government and the people. Spectrum related policies are important for the construction of smart nation.

The realization of spectrum sharing is faced with many open and critical challenges. For example, can low power transmission of secondary user meets the transmission requirements, what transmit power level should be adopted and how to control it, how to effectively estimate the users interference [33], etc. All these questions need to be addressed in the policy design.

## 2.4 India

## 2.4.1 White space sharing in India

Unlike developed countries, India has a lot of licensed but unutilized TV spectrum band [34], namely, TVWS. Since the UHF television band has ideal radio propagation characteristics, Indian authorities plan to use this spectrum band and fiber-based point of presence (POP) connections to cover the 250,000 rural offices. Since most Indian citizens live in rural areas, the communications engineers of India addressed the possibility to use TVWSs as the broadband access in rural areas [35]. Some researchers in India have proposed to create a geo-location database that can broadcast TVWS information on rate constrained channels, such that multi-hop mesh network can operate in TV UHF band [36]. In this solution, the TV UHF band is used to provide seamless connectivity between the Gram Panchayat and village users [37]. The network can coexist with TV broadcasts through a licensed shared access mechanism.

## 2.4.2 National projects of cognitive radio in India

India government has launched some projects on cognitive radio. Among them, an emergency network based on cognitive radio is proposed in adaptive ad-hoc free band (AAF) project [38]. Besides, the authorities of India have approved the spectrum sharing that allows operators to share the spectrum of a particular band to improve spectrum utilization.

## 2.4.3 Policy and regulation challenges of spectrum sharing in India

Many citizens in India cannot access to broadband Internet, hence the authorities in India are determined to change this situation. Hence spectrum sharing is a promising way to solve this problem. In additional, the government of India approved operators to share a specific spectrum band within the service area to improve spectrum utilization, which is a big step towards spectrum sharing.

## 2.5 Korea

## 2.5.1 White space sharing in Korea

The South Korea government has recently proposed a TVWS field demonstration system that meets the requirements of Federal Communications Commission (FCC). The measurements of TV band device (TVBD) networks have been conducted in Jeju, South Korea. The measurement results show that the service's coverage of mobile network can be extended to more than 2km when TVBD operates on the available TV spectrum bands [39]. The primary service in TV band is the digital television (DTV) [40]. With the DTV service, the co-channel and adjacent channel deployment scenarios of TVBD network are investigated in Korea.

## 2.5.2 Policy and regulation challenges of spectrum sharing in Korea

The government of Korea has launched some projects on cognitive radio. The service coverage measurements on TVBD networks were implemented in Jeju, Korea [39]. Besides, Korea Communications Commission (KCC) proposed a new spectrum strategy based on dynamic spectrum access technology, such as LBT universal spectrum for RFID, UWB-DAA and FACS (Flexible Access) [41]. The KCC also considers spectrum sharing for TVWS after the transition from ATV to DTV. Thus KCC pays much attention on the improvement of spectrum efficiency [41].

## 2.6 Australia

Australia is a large continental country with 23.8 million populations. However, the distribution of populations in Australia is highly inhomogeneous. Australia is highly urbanized. Nearly half of the populations live in Sydney and Melbourne. Most populations in Australia live in the big cities near the coastline. About 70% of Australia's land is arid or semi-arid. And most of the central area is not suitable for human habitation. Hence in the central area or rural area of Australia, the population is very rare and the distribution of people is very sparse.

According to Prof. Y. Jay Guo's report, the federal government of Australia has started a 43 billion dollar project to build a national broadband network (NBN) [42]. NBN will deliver 100Mbps broadband services to 90% of the population. The rest of the populations, which mainly live in rural area, will be serviced by wireless and satellite communications networks [42]. In this situation, the TV white spaces are proposed as a suitable spectrum band to provide coverage for rural area.

In Australia, with the process of switching off ATV beginning in 2010 and completing by the end of 2013, the UHF spectrum allocated to broadcasting in Australia shrinks from 300 MHz (520~820 MHz) to 174 MHz (520~694 MHz) [43]. Thus there are considerable TVWSs emerging in Australia. To provide wireless services to rural area of Australia, CSIRO is developing cognitive radio technologies aiming to dynamically access the TVWSs. These technologies include reconfigurable antennas, reconfigurable active devices, out of band interference suppression techniques, and adaptive and reconfigurable multiple input and multiple output (MIMO) systems [42]. Hence we can conclude that due to the special geographical feature and population distribution, Australia has great motivation to perform spectrum sharing. Besides, Australia has already some achievements in spectrum sharing.

# 3. Spectrum Sharing Test-bed in China

The explosive growth of mobile Internet has led to a surge of data traffic. Hence the next generation cellular network is facing both opportunity and challenge. Under this background, the new unlicensed spectrum widely known as TVWS offers new opportunities to improve spectrum efficiency. 4G cellular networks adopt long term evolution (LTE) technology based on all-IP network architecture to significantly improve area spectrum efficiency [44]. LTE can support both frequency division duplex (FDD) and time division duplex (TDD) operations over a variety of allocated frequency bands. In Asia, typically 1800, 1900, 2100, 2300 and 2500 MHz are used [45]. To meet the requirement of high data rate services, TDD LTE (TD-LTE) with flexible downlink and uplink time slot ratio without stringent requirements for paired uplink and downlink frequencies can better utilize fragmented TVWS.

However, the available licensed spectrum bands for 4G cellular networks using LTE technology in Asia are around 2 GHz and above, which increases the network cost for large area coverage. According to spectrum occupancy measurements in Beijing, significant white space opportunities exist in TV broadcasting bands around 700 MHz [46]. Hence the 4G cellular networks can be operated in TVWS. Such opportunistic operation will require dynamic spectrum access (DSA), which is based on effective cognitive radio principles. Compared to the strict requirements of the paired downlink and uplink spectrum bands with appropriate spectrum separation in FDD LTE, the TD-LTE system with multiplexed downlink and uplink in the same spectrum band can fully utilize the fragmented TVWS. Besides, to meet the service demands in hot spots, multi-tier heterogeneous radio access networks are deployed with overlapped coverage areas, which will bring challenges for spectrum allocation and interference coordination among multi-tier heterogeneous networks. Therefore DSA and efficient spectrum management technologies should be applied in heterogeneous radio access networks.

Standardization activities on TVWS utilization include IEEE 802.22 [47] for wireless regional area networks (WRAN), European Computer Manufacturers Association (ECMA) 392 [48] for wireless personal area network (WPAN) and IEEE 802.11af task group [49] for wireless local area network (WLAN). Overall, the above standards consists of the mechanisms for co-existence between secondary and primary networks, where a variety of use cases are addressed such as long-range outdoor services for 802.22 and short range indoor services for ECMA-392 and 802.11. With regards to spectrum sensing, these standards mostly advocate a quiet period for spectrum sensing by interrupting an existing communication session. This is not appropriate for TD-LTE system which has strict QoS requirements for latency in the voice service, which precludes interrupting a communication session [50]. Therefore new and simple modifications to the existing TD-LTE protocol stack are needed to exploit TVWS.

In this section, a cognitive radio enabled TD-LTE test-bed that operates in TVWS is proposed, which supports multi-tier heterogeneous radio access networks. Specifically, an efficient feature detection based spectrum sensing method is proposed for TD-LTE which can achieve 99.9% detection probability and 1% false alarm probability. The proposed cognitive radio enabled TD-LTE frame structure utilizes the guard period and adjacent vacant uplink subframe for spectrum sensing by the Cognitive eNodeB (CeNB), which requires minimum protocol stack modifications. In particular, Cognitive Control Channel (CogCCH) is proposed in medium access control (MAC) layer to transmit CR-related Radio Resource Control (RRC) information. Physical Cognitive Channel (PCogCH) is proposed in Physical (PHY) layer to broadcast the spectrum decision information to CR user. To mitigate any potential adjacent channel interference between the TD-LTE system and the TV broadcasting system, appropriate guard bands between these two systems are needed. The width of guard bands is determined through simulations and field experiments. To the best of our knowledge, the proposed test-bed is the first CR enabled TD-LTE system operating in TVWS.

## 3.1 CR enabled TD-LTE test-bed utilizing TVWS

The TD-LTE based spectrum sharing test-bed is developed by Beijing University of Posts and Telecommunications (BUPT). To solve the challenges of deploying CR enabled TD-LTE system operating in TVWS, both solution and analysis are proposed, including an efficient spectrum sensing method for TV signal detection, interference analysis and the CR enabled TD-LTE protocol stack to support the efficient utilization of TVWS, which are introduced in the following sections.

### 3.1.1 Spectrum sensing in TVWS

The Cognitive eNodeB (CeNB) based spectrum sensing method is proposed for spectrum sensing in TVWS, which has two advantages. Firstly, the spectrum occupied and signaling overhead for information exchange between CR users and CeNBs can be minimized. Secondly, the energy consumption for spectrum sensing by CR users can be saved.

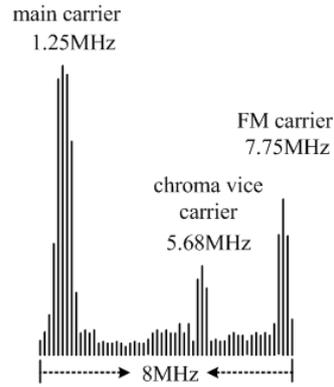

Fig. 11 Spectrum characteristics of analog TV signals

Unlike the US where TVWS exists among digital TV signals, analog TV signals will be the dominant primary users in China for many years. Therefore, a feature detection based spectrum sensing method exploring the specific characteristic of the analog TV signals is proposed. According to the spectrum characteristics of analog TV signals (PAL-D) in China [7], the energy of the baseband TV signals is mostly concentrated on 1.25 MHz, chroma vice carrier on 5.68 MHz and audio FM carrier on 7.75 MHz as illustrated in Fig. 11. The procedure of the proposed feature detection method for analog TV signals is shown in Fig. 12. First, the TV RF signal is digitized directly via fast ADC, which is then down-converted to a baseband signal. If there are narrow band signals existing on 1.25 MHz, 5.68 MHz and 7.75 MHz, and the signal waveform is similar to the analog TV signals, then the analog TV signal is detected. Otherwise, the spectrum is assumed not to be occupied by TV signal. Because the detection bandwidth is only 200 kHz compared to the 8 MHz for each TV channel, it can increase the spectrum sensing speed for each TV channel with an improvement of 16 dB. The proposed feature detection method can also reduce the noise and the sampling time, which has a good performance under the low signal to noise ratio (SNR) condition.

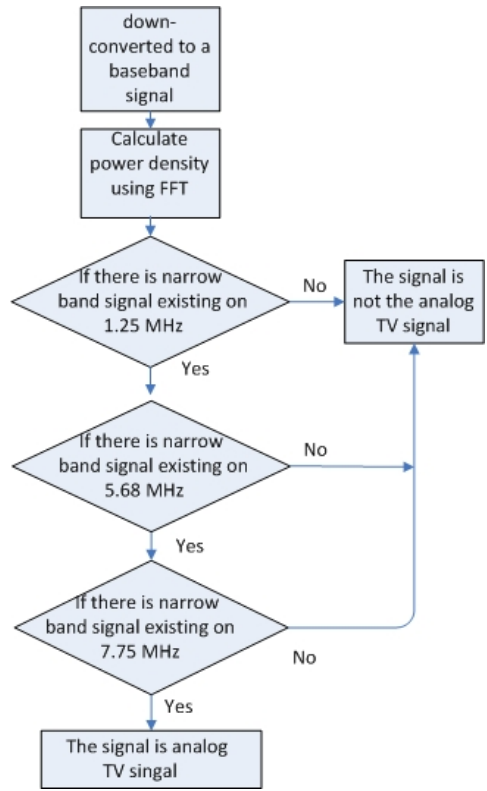

Fig. 12 Procedure of the feature detection method for analog TV signals

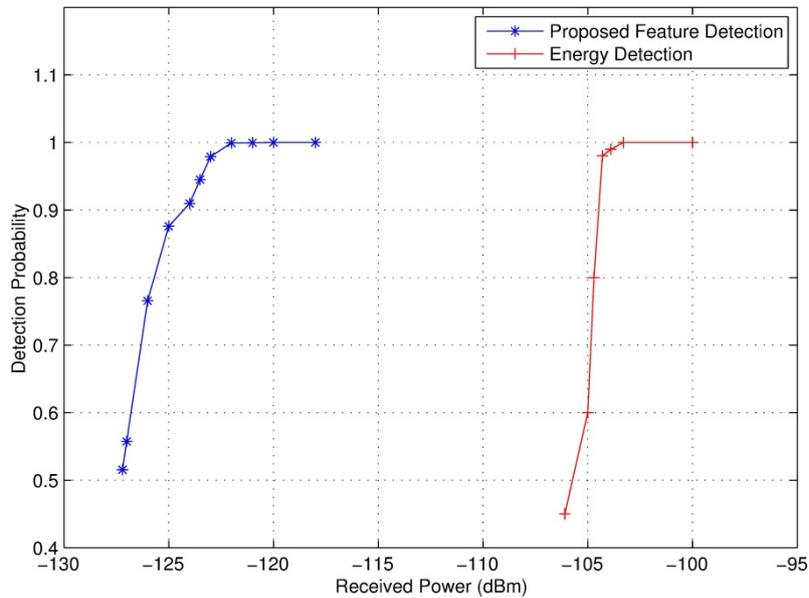

Fig. 13 Detection probability against the received power level of TV signals

The field experiments are implemented to test the performance of the proposed feature detection method, which is programmed into a baseband signal process board of the spectrum sensing modules. The Agilent N5182A MXG signal generator is used to generate analog TV signals. Spectrum sensing module receives analog TV signals through a cable and reports the detection results to the computer. The detection probability and false alarm probability can be determined by the experiment in Fig. 13, which shows that the detection probability can be 99.9% under 1% false alarm probability. Besides, the proposed feature detection method improves the weak signal detection capability significantly with the TV signal power level of -120 dBm.

Furthermore, in order to utilize the vacant spectrum in time-frequency-space multi-domains in TVWS, an efficient spectrum utilization scheme called three regions is proposed in [53], which applies the joint geo-location database and spectrum sensing scheme in order to reduce the complexity and cost of traditional spectrum sensing. First, CeNBs of the TD-LTE system access to the geo-location database to find their locations and decide their specific regions, i.e., the black region, grey region and white region. Geo-location database is a database system that maintains records of all authorized services in TV frequency bands, and is capable of determining vacant channels at a specific location based on the interference protection requirements. Black region is defined as a region where the CeNB within the black region cannot reuse the spectrum of TV broadcasting system. Grey region is defined as a region where the CeNB should apply the spectrum sensing to utilize TVWS. White region is defined as a region where the CeNB can reuse the spectrum of the TV broadcasting system freely without implementing the spectrum sensing procedure. Therefore by checking the position information in the geo-location database, the CeNB can decide its specific region and make an appropriate decision on whether to apply the spectrum sensing or not. The scheme of three regions can improve the efficiency and reduce the cost of spectrum utilization.

## 3.1.2 Interference analysis

One approach to reduce the adjacent channel interference (ACI) between TD-LTE system and TV broadcasting system is reconfiguring the central frequency of the TD-LTE system and setting appropriate guard bands between two systems, which will reduce available spectrum resources. Another approach is to use power control scheme to transmit with lower power on adjacent channels closed to the active TV channels. Based on the geo-database approach, if the TD-LTE system is close to TV receivers, it can use the spectrum bands far from the active TV channels to reduce the interference.

Firstly, we conduct simulations to analyze the interference between two systems using Matlab. We establish the TV circle-shaped system and TD-LTE honeycomb-shaped network topology (19 CeNBs and 57 sectors). Then we assign TV receivers and TD-LTE users, configure transmitter/receiver and channel models, upload resource scheduling and user mobility schemes to simulate the real system operation states. Moreover, TV receivers and TD-LTE users are supposed to be uniformly distributed. During the simulations, system configurations are the category B in [54]. The left part of Fig. 14 shows the interference probability of the TV broadcasting system versus ACIR, where interference probability is defined as the percentage of TV users that cannot receive signals normally due to the interference. It denotes that the interference decreases as ACIR becomes larger. It also depicts that the interference from CeNB is bigger than that from CR users, which is caused by the much higher CeNB transmitter power compared to CR user. The right part of Fig. 14 shows the interference from the TV broadcasting system to the TD-LTE system. Similarly, we define the percentage of interference-caused capacity decline as the capacity loss and the larger ACIR leads to less capacity loss of the TD-LTE system and the uplink capacity is prone to be decreased compared to the downlink capacity. In general, from the engineering point of view, 5% system performance loss is acceptable for the TD-LTE system. Thus, 75/30dB ACIR (TV interfered by TD-LTE downlink/uplink) is acceptable for the TV broadcasting system and 27/78dB ACIR (TD-LTE downlink/uplink interfered by TV) is acceptable for the TD-LTE system.

Based on the mapping relationship between ACIR and the width of guard bands [54], the frequency separation bandwidth of 7 MHz is appropriate to enable the coexistence of two systems.

Secondly, further field experiments are performed to find out the width of the guard bands. The Agilent E4438C signal generator and Agilent N5182A MXG signal generator are applied to generate downlink TD-LTE and TV broadcasting signals, respectively, which go through the signal emulator (EB propsim F8) to emulate different indoor and outdoor channel environments. Then, an R&S ETL TV broadcasting signal analyzer acts as a TV receiver to measure the quality of received TV signals with the interference from TD-LTE signals. The central frequency of both TD-LTE and the TV signals is set to 700 MHz at first, and then the central frequency of the TD-LTE system is increased until the TV signal analyzer indicates that the received TV signal quality reaches the normal standard [52]. Then, the separation bandwidth between two signals is the width of the guard bands. The results show that a frequency separation of 6 MHz between TD-LTE and TV signals is an appropriate value, which will not cause harmful interference between them. Therefore, the 7 MHz guard band can be used for the coexistence between two systems.

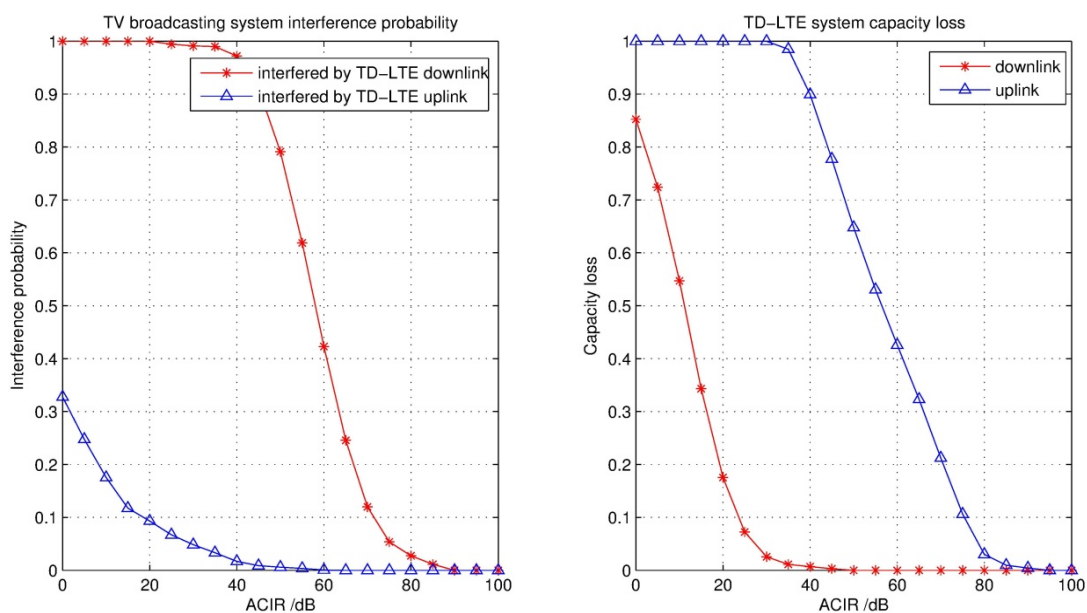

Fig. 14. Performance of the TV broadcasting system and the TD-LTE system against ACIR

## 3.1.3 The CR enabled TD-LTE protocol stack

The protocol stack of the TD-LTE system should be designed with the minimum modifications to support CR functions. Besides, the spectrum sensing results and network information exchange among CeNBs should be supported. Therefore, modifications to the TD-LTE protocol stack for both the CeNB and the CR user are designed to support CR functions, as illustrated in Fig. 15.

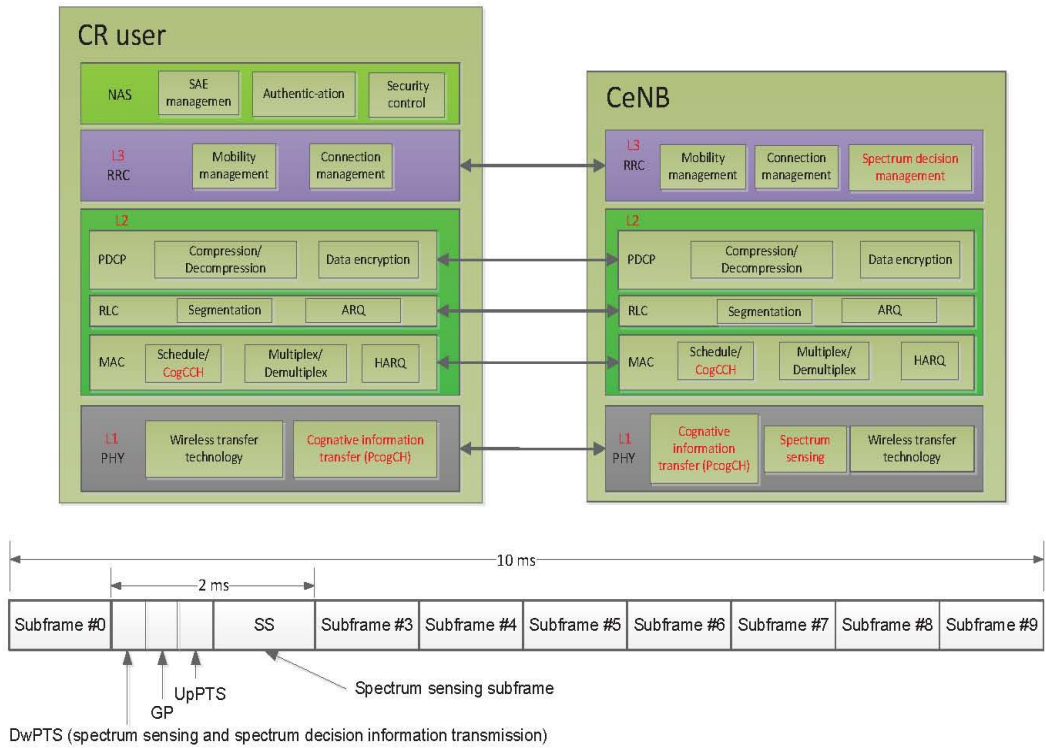

Fig. 15. Proposed CR enabled TD-LTE protocol stack and frame structure

In the RRC layer, we introduce the spectrum decision management module in CeNB to make spectrum decision according to the spectrum sensing information received from the spectrum sensing module. Two main spectrum management functions are proposed for RRC. The first one is called the long-term spectrum management among different CeNBs for the efficient allocation and utilization of TVWS. The second one is called the short-term spectrum management within CeNB.

The MAC layer provides different types of data transmission on different types of logic channels, so the types of logic channels are determined by data types. To accomplish the interaction of CR-related information, a new logic channel called CogCCH is proposed both in CeNB and CR users to transmit CR-related RRC messages.

In the PHY layer, a physical channel called PCogCH is proposed both in CeNB and CR user to carry the spectrum decision information on the transport channel called Cognitive Channel (CogCH) and broadcast the spectrum decision information to CR users through the special subframe. The CR user executes the spectrum decision after receiving the broadcast information. To reduce the power consumption and signaling overhead for the CR user, the spectrum sensing module is only added in CeNB to perform the spectrum sensing and report the spectrum sensing results to RRC layer. Meanwhile, through the X2 interface among CeNBs, cooperative spectrum sensing can be performed accordingly, which will be used between the adjacent CeNBs for spectrum sensing information exchange and long term spectrum management.

Furthermore, the TD-LTE frame structure is modified to enable CR functions as illustrated in the lower part of Fig. 15. Within 10 ms TD-LTE frame, the Guard Period (GP) in the special subframe #1 is used for spectrum sensing which will not disrupt the communication between CeNBs and CR users. Moreover, the adjacent vacant uplink subframe #2 is also used for spectrum sensing when the candidate TVWS is too wide. Thus, the TD-LTE system performs spectrum sensing within 2 ms in each 10 ms frame period. Meanwhile, the spectrum sensing module will perform spectrum sensing and deliver the results to the RRC layer. The DwPTS in the special subframe is chosen to transfer both spectrum sensing and spectrum decision information of the previous frame.

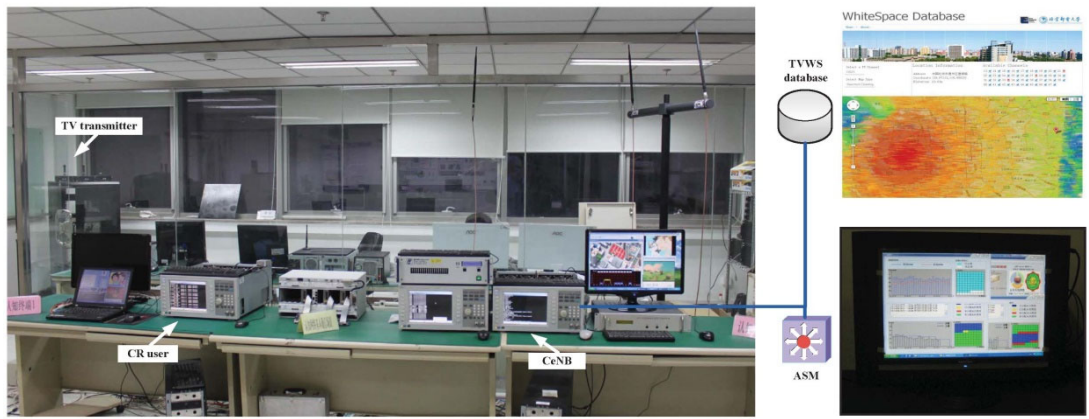

Fig. 16 CR enabled TD-LTE Test-bed

## 3.2 Test-bed and results

Considering the industrial trend of future telecommunication networks in China, a CR enabled TD-LTE test-bed has been designed and developed in Beijing University of Posts and Telecommunications (BUPT) with two CeNBs and eight CR users. Both CeNBs and CR users are implemented with a unified hardware platform [55], where the baseband signal processing functions are implemented on TI C6487 DSP. In the three-core DSP, one of the DSP core is used for downlink signal processing, the second one is used for uplink signal processing and the third one is for scheduling. AD/DA operation is implemented in Xilinx FPGA, with the AURORA interface to DSP. Besides, the platform can also download different protocols from DSP dynamically, which is capable of reconfiguring its work mode and parameters intelligently to efficiently utilize TVWS.

Each TV channel occupies a frequency of 8 MHz in China, thus the TD-LTE system can be implemented with 20 MHz system bandwidth when three continuous TV channels (24 MHz) are vacant. After the CeNB detects the TV signal, the CeNB will execute spectrum handover and switch to another vacant spectrum band accordingly. Furthermore, the spectrum sensing results will also be sent to the TVWS database for vacant spectrum information update, which is used for vacant spectrum information coordination and synchronization among CeNBs. In addition, considering the scenario that continuous TV channels are less than three, the test-bed can apply the dynamic system bandwidth adjustment technology, which means that 15 MHz bandwidth is used when two continuous TV channels are vacant and 5 MHz bandwidth is used if only one TV channel is available. Therefore, the test-bed can dynamically change its system bandwidth in order to adapt to different vacant spectrum conditions in practical scenarios.

Proposed CR enabled TD-LTE test-bed has both the outdoor and indoor scenarios. In the outdoor scenario, CeNB is deployed on the roof of the building along the XINGTAN road with two mobile CR users. In the indoor scenario, the trial of the test-bed is illustrated in Fig. 16 with three analog TV signal transmitters deployed as the primary TV broadcasting system. The CeNB and CR users are placed inside the room with a Light-of-Sight (LoS) propagation. The antennas for both CeNB and the CR users are 6 dBi rod-antenna. The transmit power of the CeNB is 20 dBm. In addition, the CeNBs are connected to the TVWS database and advanced spectrum management (ASM) subsystem. The TVWS database will collect and update the crude data of spectrum sensing results from different CeNBs, the database of mobile network, TV broadcasting operators and the database of state spectrum regulators, in order to draw the global spectrum occupancy graph. Moreover, based on the global spectrum occupancy graph from the TVWS database, the ASM subsystem is responsible for an efficient spectrum management and coordination among different CeNBs by making decisions for the efficient vacant spectrum allocation.

In Fig. 17, the packet loss ratio of the CR enabled TD-LTE system is plotted during a period of 2 s by 200 samples with 10 ms for each sample. During the experiment, the TV transmitter is turned on at 1 s, and the packet loss ratio of the CR enabled TD-LTE system is fairly steady before 1 s and increases significantly after 1 s as depicted by the surge at the point of 100 samples. This indicates that the interference from TV signals is strong, which results in a tremendous packet loss. However, due to the accurate and timely TV signal detection and spectrum handover, the packet loss ratio of the CR based TD-LTE system can be quickly restored within a very short time of 50 ms as shown in Fig. 5. In contrast to the silence period applied by IEEE 802.22 standard in [2], the proposed CR enabled TD-LTE test-bed can achieve the spectrum handover within 50 ms on average without severe performance deterioration or service dropout, which significantly outperforms the avoiding time of 2 s in [2].

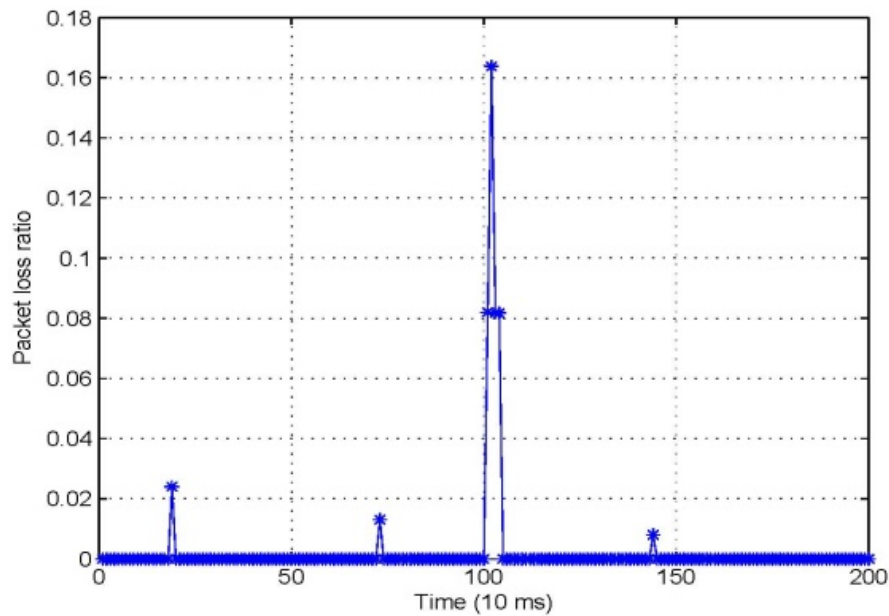

Fig. 17 Packet loss ratio during spectrum handover